\documentclass[useAMS,usenatbib]{mn2e}

\usepackage{color}
\usepackage{txfonts}
\usepackage{dcolumn}
\usepackage[dvipdfm]{graphicx}

\title[PMR of Mira-like variables in the LMC as a tool to understand circumstellar extinction]{Period-Magnitude relation of Mira-like variables in the Large Magellanic Cloud as a tool to understand circumstellar extinction}
\author[Yoshifusa Ita \& Noriyuki Matsunaga]{Yoshifusa Ita$^{1}$\thanks{E-mail:
yita@astr.tohoku.ac.jp}, and Noriyuki Matsunaga$^{2,3}$\\ 
$^{1}$Astronomical Institute, Graduate School of Science, Tohoku University, 6-3 Aramaki Aoba, Aoba-ku, Sendai, Miyagi 980-8578, Japan\\
$^{2}$Institute of Astronomy, School of Science, The University of Tokyo, Mitaka, Tokyo 181-0015, Japan\\
$^{3}$Kiso Observatory, School of Science, The University of Tokyo, Mitake, Kiso, Nagano 397-0101, Japan\\
}
\begin{document}

\date{Accepted 1988 December 15. Received 1988 December 14; in original form 1988 October 11}

\pagerange{\pageref{firstpage}--\pageref{lastpage}} \pubyear{2002}

\maketitle

\label{firstpage}

\begin{abstract}Near- to mid-infrared period-magnitude relations and also the period-bolometric luminosity relation of OGLE-III Mira-like variables in the LMC are derived. The relations have a kink, and the period at which the break occurs is quantitatively obtained. There are many Mira-like variables whose fluxes at the optical and the near-infrared wavebands are fainter than the ones predicted by the period-magnitude relations. The deviation is due to the circumstellar extinction, and the amount of the deviation is found to be strongly correlated with near-infrared colors. The empirical formulae relating the amount of the deviation and the near-infrared colors are derived. These relations are useful to accurately calculate the distances to the dusty Mira-like variables, because the dimmed fluxes due to the circumstellar extinction can be estimated. In a manner analogous to the interstellar extinction law, the ratios of deviations at any two different wavebands are calculated. The ratios are found to change with the pulsation period, indicating that the dust properties are subject to change as Mira-like variables evolve.
\end{abstract}


\section{Introduction}
One of the most astronomically profitable aspects to study radially pulsating variable stars is that some groups of them have their individual period-luminosity relations that can be used as cosmic distance scales. Mira-like variables are one of such group of variable stars that evolved from low- to intermediate mass ($\sim0.8 < M/M_\odot < \sim8$) stars, and they are in the late stage of stellar evolution (e.g., \citealt{iben}). Since the discovery of the period-magnitude relation of Mira-like variables (\citealt{glass1981}), many astronomers observed them to refine the relation and to study its metallicity dependency. From late 90's to early 00's, there was a major progress on the study of the period-luminosity relations of variable stars due to the advent of the optical large survey projects looking for the gravitational lensing events with robotic telescopes (e.g., MACHO, OGLE, EROS, MOA). The discovery of parallel sequences of period-magnitude relations of red giant variables in the Large Magellanic Cloud (\citealt{wood2000}) is an especially important result, because it told us that each group of variables has different period-luminosity relations, and also, different relation is assigned to different pulsation mode within a group (see also, for example, \citealt{kiss2003}, \citealt{ita2004}, \citealt{groenewegen2004}, \citealt{fraser2005}, \citealt{derekas2006}, \citealt{glass2009}, \citealt{soszynski2009}, and \citealt{tabur2010}). 

The aim of this paper is to study the near- to mid-infrared period-magnitude relations of Mira-like variables in the LMC, using the combined data of recently published OGLE-III data (\citealt{soszynski2009}) and the \textit{Spitzer} SAGE catalog (\citealt{meixner2006}). Infrared data is imperative because Mira-like variables are usually associated with the circumstellar dust shell, which is formed by the mass-loss phenomenon.

\section{The data}
Recently, the OGLE project \citep{udalski2003} released its third phase (OGLE-III; 12th June, 2001 -- 3rd May, 2009) survey data. They provided catalogs of several types of variable stars in the Magellanic Clouds. Using their on-line database\footnote{visit the OGLE web page at http://ogle.astrouw.edu.pl/}, stars with variable star type of "Mira" were chosen by querying their catalog of long-period variables in the LMC \citep{soszynski2009}. The query yielded a catalog of 1,663 Mira-like variables in the LMC. \citet{soszynski2009} used the $I-$band pulsation amplitude to distinguish Miras and Semi-regular variables. They defined Miras as having $I-$band pulsation amplitude greater than 0.8 mag. In this paper, we follow their classification and call these Mira-like variables as just Miras or OGLE-III Miras. The OGLE-III catalog lists not only coordinates, but also pulsation periods, amplitudes, surface chemistries (Oxygen-rich / Carbon-rich) inferred by photometric colors, time averaged $V-$ and $I-$band magnitudes, and other useful parameters.

\subsection{Cross-identification with existing catalogs}
The OGLE-III Mira catalog is cross-identified with the following existing catalogs using a positional tolerance of 3 arcsec. If more than one stars are present within the tolerance radius, the closest one is adopted and discard the others. The result is used for discussion in the rest of this paper. 

\begin{itemize}
\item The Large Magellanic Cloud Photometric Survey catalog \citep{zaritsky2004}: The catalog lists $U$, $B$, $V$, and $I$ stellar photometry of the central 64 deg$^2$ area of the LMC. It must be noted that we preferentially used the time-averaged $V-$ and $I-$band data in the OGLE-III survey catalog whenever available. 1,412 out of a total of 1,663 Miras have counterparts in this catalog.
\item The Two Micron All Sky Survey (2MASS) catalog \citep{skrutskie2006}: The catalog provides uniform $J$, $H$, and $K_s$ photometry for sources all over the sky. The 2MASS catalog is complete down to $K_s < 14.3$ mag in the absence of confusion\footnote{http://www.ipac.caltech.edu/2mass/releases/allsky/doc/sec2\_2.html}. The 2MASS magnitudes are used for the bright sources without IRSF measurements (The IRSF survey did not detect bright sources due to the saturation limit of about 10 mag at $K_s$ band. See below). 1,639 out of a total of 1,663 Miras have counterparts in this catalog.
\item The IRSF Magellanic Clouds Point Source Catalog (IRSF catalog; \citealt{kato2007}): The IRSF catalog lists $J, H,$ and $K_s$ photometry of over 1.4$\times$10$^7$ sources in the central 40 deg$^2$ area of the LMC. Compared to the contemporary DENIS (\citealt{cioni2000a}) and 2MASS (\citealt{skrutskie2006}) catalogs, the IRSF catalog is more than two magnitudes deeper at $K_s$ band and about four times finer in spatial resolution. The IRSF system magnitudes were converted into the 2MASS system ones by using the conversion equations given in \citet{kato2007} and \citet{kucinskas2008}. We preferentially used IRSF photometry whenever available. 1,557 out of a total of 1,663 Miras have counterparts in this catalog.
\item The \textit{Spitzer} SAGE-LMC survey catalog \citep{meixner2006}: The catalog lists near- ([3.6], [4.5], [5.8], and [8.0]) to mid-infrared ([24] and [70]) photometry of sources in the central 49 deg$^2$ area of the LMC. Throughout this paper, the numbers bracketed by $[~]$ designates the data of the \textit{Spitzer} catalog, for example, [3.6] indicates the photometry in the 3.6 $\mu$m band. The SAGE-LMC team recently made a final data delivery (DR3)\footnote{visit http://ssc.spitzer.caltech.edu/legacy/sagehistory.html}. They provided two types of the catalog. One is the "Catalog", which is a more highly reliable list. The other is the "Archive/Full", which is a more complete list. See the explanatory document "The SAGE Data Products Description" prepared by Dr. M. Meixner for more detailed descriptions of the catalogs. We use the "Archive/Full" version for the cross-identification. 1,612 and 1,274 out of a total of 1,663 Miras have counterparts in the near- and mid-infrared SAGE-LMC catalog, respectively.
\item Optical carbon star catalog \citep{kontizas2001}: The catalog lists 7,760 carbon stars in the LMC that were identified by the optical objective-prism spectroscopy survey. 163 out of a total of 1,663 Miras have counterparts in this catalog.
\item \textit{Spitzer} IRS-identified O-rich and C-rich evolved stars \citep{groenewegen2009}: \citet{groenewegen2009} uniformly reduced all \textit{Spitzer} IRS data of evolved stars in the Magellanic Clouds taken so far. They calculated luminosities and mass-loss rates for 66 O-rich and 68 C-rich infrared stars in the LMC. 40 out of a total of 1,663 Miras have counterparts in their list.
\end{itemize}

\subsection{Corrections for the interstellar reddening}
In this paper, the optical ($U, B, V,$ and $I$) and the near-infrared ($J, H,$ and $K_s$) photometry are corrected for the interstellar reddening based on the relations in \citet{cardelli1989}, assuming $A_V/E(B-V) =$ 3.2. We uniformly adopt ($A_U$, $A_B$, $A_V$, $A_I$, $A_J$, $A_H$, $A_{K_s}$) $=$ (0.407, 0.345, 0.272, 0.159, 0.078, 0.048, 0.032) mag, corresponding to the mean foreground reddening of $E(B-V) =$ 0.085 mag toward the LMC (\citealt{larsen2000}). Fluxes in any other longer wavelengths (i.e., photometry in the Spitzer wavebands) are not corrected for the interstellar extinction, which we assume negligible. We also ignore the reddening inside the LMC.

\subsection{Calculation of the bolometric luminosity}
We calculate the bolometric luminosity of all OGLE-III Miras by using a cubic spline to interpolate the spectral energy distribution and integrate it from the shortest available wavelength to 8 or 24 $\mu$m. We used zero-magnitude fluxes tabulated in Table~\ref{table:zeromag} to convert magnitude into Jansky. Color-correction is not applied to the flux density, due to the lack of information of the incident spectrum. For some of the very red sources, the calculated luminosities can be underestimated to a large extent because the fluxes longward of 24 $\mu$m are not included. Therefore the calculated luminosities should be only lower limits for the very red sources. Also, we have to note that the calculated luminosities are rather uncertain, because they are sensitive to changes of degree of interstellar reddening, color correction, and also to time variation of light, which is significant especially for Mira-like variables discussed here.

\newcolumntype{d}[1]{D{.}{\cdot}{#1}}
\newcolumntype{.}{D{.}{.}{-1}}
\newcolumntype{,}{D{,}{,}{-1}}
\begin{table}
  \caption{The adopted zero magnitude flux densities and their corresponding reference wavelengths.}\label{table:zeromag}
  \begin{center}
    \begin{tabular}{c D..{2} D..{4} r}
    \hline
    Wavebands & \multicolumn{1}{c}{$f_0$$^1$} & \multicolumn{1}{c}{$\lambda_r$$^2$} & Ref.$^3$ \\
              & \multicolumn{1}{c}{[Jy]} & \multicolumn{1}{c}{[$\mu$m]} &  \\
    \hline
    $U$ & 1649 & 0.3745 & a \\
    $B$ & 4060 & 0.4481 & a \\
    $V$ & 3723 & 0.5423 & a \\
    $I$ & 2459 & 0.8071 & a \\
    $J$ & 1594 & 1.235  & b \\
    $H$ & 1024 & 1.662  & b \\
    $K_s$   & 666.8 & 2.159 & b  \\
    $[3.6]$ & 280.9 & 3.550   & c \\
    $[4.5]$ & 179.7 & 4.493   & c \\
    $[5.8]$ & 115.0 & 5.731   & c \\
    $[8.0]$ & 64.1  & 7.872   & c \\
    $[24]$  & 7.14  & 23.68   & d \\
    \hline
    \end{tabular}
  \end{center}
$~^1$ Zero magnitude flux density, $~^2$ Reference wavelength, $~^3$ {\bf References :}
$a$ \citet{cohen2003a}; $b$ \citet{cohen2003b}; $c$ \citet{irac2006}; $d$ \citet{mips2008}
\end{table}

\begin{figure}
\centering
\includegraphics[scale=0.46,angle=-90]{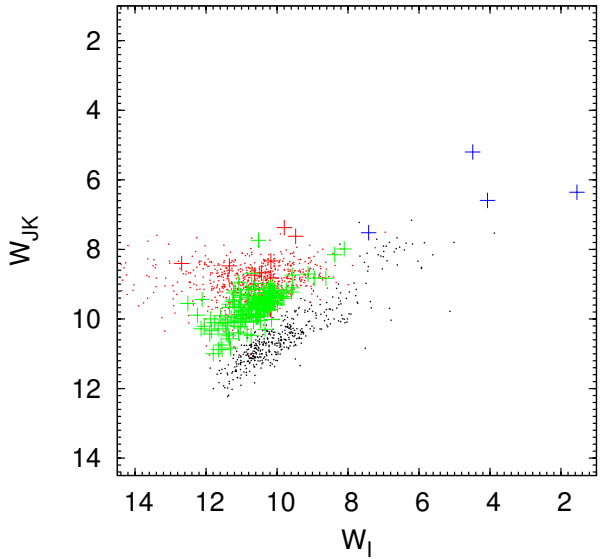}
\caption{The $W_I$ vs. $W_{JK}$ diagram of OGLE-III Miras in the LMC, where $W_I$ and $W_{JK}$ are Wesenheit indices (see text). The black and red small dots show the color-classified O-rich and C-rich chemistries (\citealt{soszynski2009}), respectively. Miras with spectroscopically known surface chemistry are shown in big pluses. The green and red pluses show optical and infrared C-rich stars identified by \citet{kontizas2001} and \citet{groenewegen2009}, respectively. The blue pluses indicate O-rich red supergiants identified by \citet{groenewegen2009}.}
\label{wesenheit}
\end{figure}

\subsection{Evaluation of the color-classified surface chemistry in the OGLE-III catalog}
\citet{soszynski2009} used $W_I$ vs. $W_{JK}$ diagram to classify O-rich and C-rich surface chemistry of Mira variables in the LMC, where $W_I$ and $W_{JK}$ are Wesenheit indices defined as $W_{I} = I - 1.55(V-I)$ and $W_{JK} = K_s - 0.686(J-K_s)$, respectively. The validity of their classification should be checked before use, and its evaluation is done in the following way. We make the same $W_I$ vs. $W_{JK}$ diagram as in \citet{soszynski2009}, using only 1,663 OGLE-III Miras. The diagram is shown in figure~\ref{wesenheit}. Then, those with spectroscopically known surface chemistry (based on \citealt{kontizas2001} and \citealt{groenewegen2009}) are highlighted to see their distribution on the employed Wesenheit index plane. It is clear that the distribution of Miras with spectroscopically known chemistry is fairly reproduced by the OGLE's color-classification criteria, that were defined by \citet{soszynski2009}. We do not see any color-classified O-rich stars that correspond to the infrared O-rich red supergiants (blue pluses in Figure~\ref{wesenheit}). This is due to the lack of their $V-$band data, probably because of the circumstellar extinction. Presumably, they get fainter than the detection limit of the OGLE-III $V-$band observations. We use their color-classified chemistry throughout this paper unless otherwise noticed.

\section{Results and Discussions}
\begin{figure*}
\centering
\includegraphics[scale=0.75,angle=-90]{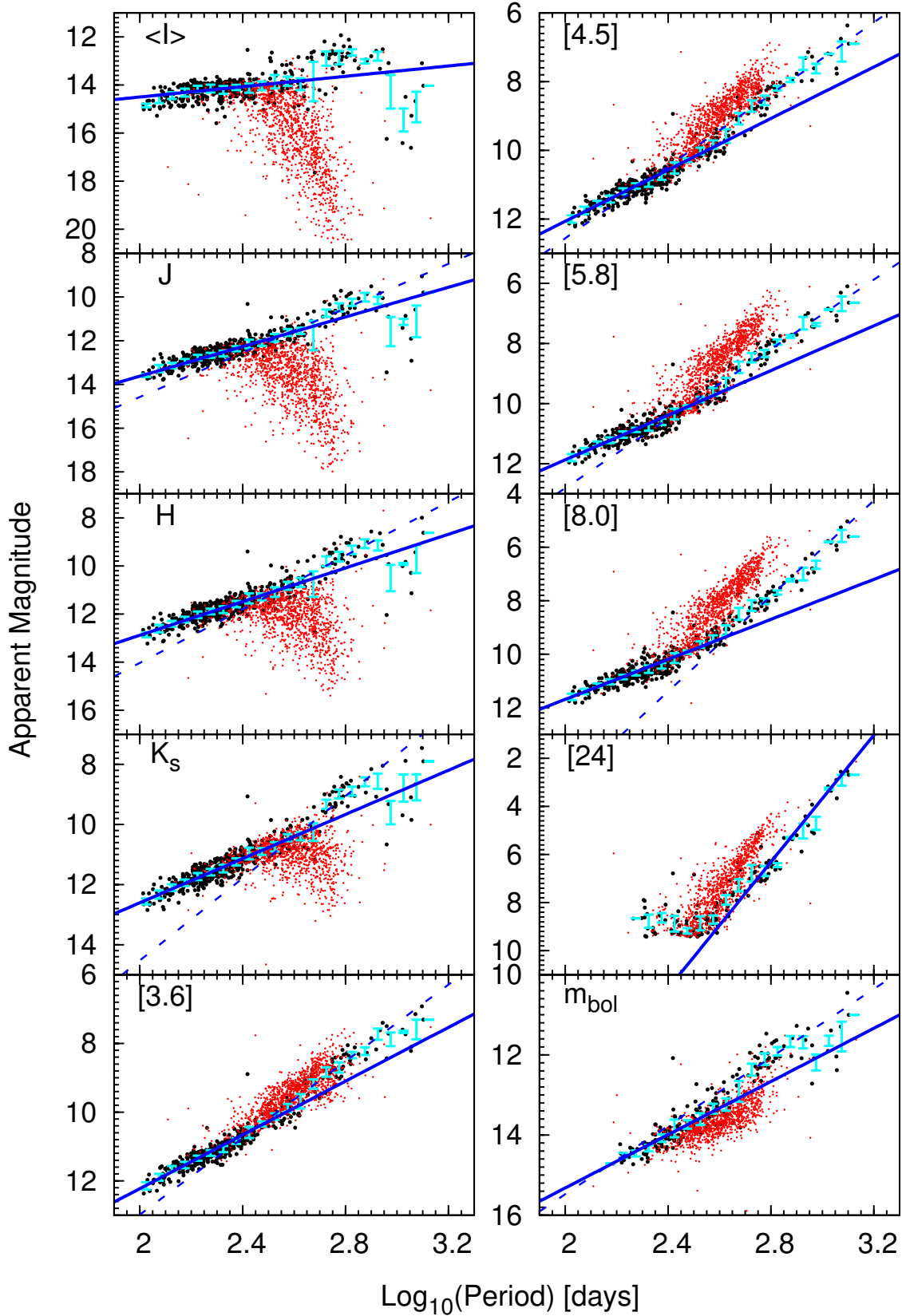}
\caption{Period-magnitude relations for OGLE-III Miras in the LMC. The color of the marks represents the differences in the color-classified surface chemistries (\citealt{soszynski2009}); black for O-rich and red for C-rich stars. The cyan vertical lines show the mean magnitudes of O-rich Miras in 0.05 mag bins and their corresponding standard deviations. The thick solid and thick dashed blue lines are least-square fit to the mean magnitudes (see text and table~\ref{plrtable}).}
\label{pl}
\end{figure*}

\begin{figure}
\centering
\includegraphics[scale=0.40,angle=-90]{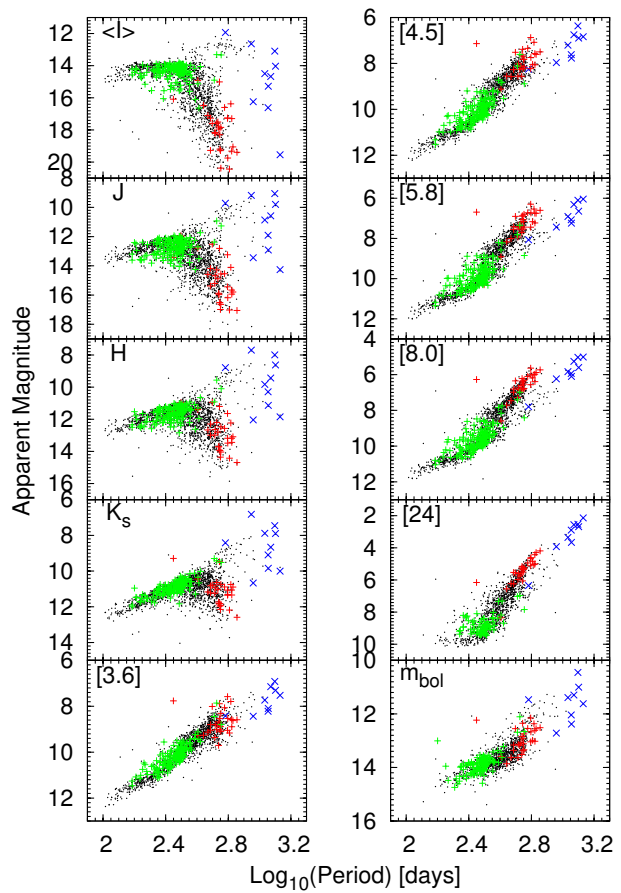}
\caption{The same as the Figure~\ref{pl}, but stars with spectroscopically known surface chemistry are highlighted. The pluses show optical (green) and infrared (red) C-rich stars identified by \citet{kontizas2001} and \citet{groenewegen2009}, respectively. The blue crosses indicates O-rich red supergiants identified by \citet{groenewegen2009}.}
\label{pl2}
\end{figure}

\subsection{Period-magnitude relations}
\citet{glass2009} used the combined data of the MACHO survey (e.g., \citealt{alcock}) and the previous version of the Spizer SAGE catalog (\citealt{meixner2006}) to discuss the mid-infrared period-magnitude relations (PMR) of variable red giants. They compared the PMRs of variable red giants in the LMC and the NGC6522, showing that there is little difference in their PMRs despite the differences in ages and metallicities between the two galaxies. Very recently, \citet{riebel} discussed the infrared PLRs and showed that the wavelength dependence of the slope of the period-luminosity relationship is different for different classes of variable red giants.

Here we use the combined data of the OGLE-III Mira catalog and the new version of the Spitzer SAGE catalog (DR3), focusing on the PMRs of Miras in the LMC. In Figure~\ref{pl}, the PMRs of the OGLE-III Miras are shown. The abscissa is the common logarithm of the primary pulsation period in days. We use the primary period only. Note that the OGLE-III catalog provides not only the primary period, but also the secondary, and the tertiary periods. The left ordinate of each panel is in reddening-corrected apparent magnitude. Its corresponding waveband is indicated at the top left of each panel. The unit of the left ordinate can be scaled to the absolute one by subtracting a certain constant (i.e., distance modulus of the LMC). The colors of the marks in Figure~\ref{pl} represent the difference in the color-classified surface chemistry, as being black for O-rich, and red for C-rich Miras, respectively. 
Figure~\ref{pl2} is the same diagram as the Figure~\ref{pl}, but highlights Miras with spectroscopically known surface chemistry. The pluses show optical (green) and infrared (red) C-rich stars identified by \citet{kontizas2001} and \citet{groenewegen2009}, respectively. The blue crosses indicates O-rich red supergiants identified by \citet{groenewegen2009}. By comparing these two figures, one can understand where each of the three groups is located in the period-magnitude plane. 

There are several interesting features in Figures~\ref{pl} and \ref{pl2}. It seems that Miras with shorter primary pulsation periods fall on a relatively tight PMR regardless of their surface chemistry. We conducted a two-tailed Student's t-test to test a hypothesis that "the O-rich and C-rich Miras obey the same period-luminosity relation". A 95 \% confidence level was chosen. Specifically, the mean magnitudes and variances of O-rich and C-rich Miras in the $\log_{10} P (days) = 0.05$ bins are calculated and then compared. The results are summarized in Table~\ref{t-test}. The test suggests that Miras with periods shorter than about 200 days ($\log_{10} P \sim $ 2.3) obey the same PMR regardless of their surface chemistry and also of employed waveband. For the easy explanation, we call the PMRs of these shorter period Miras as standard PMRs, and use them for comparison.

\begin{table}
  \caption{Results of Student's t-test to test a hypothesis that "the O-rich and C-rich Miras obey the same period-luminosity relation". The A and R means accept and reject the hypothesis, respectively.}\label{t-test}
  \begin{center}
    \begin{tabular}{ccccccccccc}
    \hline
     \multicolumn{1}{c}{$\log_{10} P$} & \multicolumn{10}{c}{Result} \\
      & $I$ & $J$ & $H$ & $K_s$ & $3.6$ & $4.5$ & $5.8$ & $8.0$ & $24$ & m$_{\textrm{bol}}$\\
    \hline
2.125 & A & - & - & - & A & A & A & A & - & - \\ 
2.175 & A & A & A & A & A & A & A & A & - & - \\
2.225 & A & A & A & A & A & A & A & A & - & A \\
2.275 & A & A & A & A & R & A & R & A & - & A \\
2.325 & R & R & A & A & R & R & R & R & A & A \\
2.375 & R & R & R & A & R & R & R & R & A & A \\
2.425 & R & R & R & A & R & R & R & R & A & R \\
2.475 & R & R & R & A & R & R & R & R & R & R \\
2.525 & R & R & R & A & R & R & R & R & R & R \\
2.575 & R & R & R & R & R & R & R & R & R & R \\
2.625 & R & R & R & R & R & R & R & R & R & R \\
2.675 & R & R & R & R & R & R & R & R & R & R \\
    \hline
    \end{tabular}
  \end{center}
\end{table}

The PMRs become complex for Miras with pulsation periods longer than about 320 days. C-rich Miras with longer periods locate below the extension of the standard PMRs in the optical and near-infrared wavebands, although they locate above the standard PMRs in the \textit{Spitzer} wavebands. This can be understood as the extinction by the circumstellar dust, and also as the intrinsic redness of C-rich Miras. On the other hand, O-rich Miras with longer periods locate above the reference PMR in all wavebands. Some O-rich Miras with very long period ($\log_{10}$ P $>$ 3.0) locate below the reference PMR, due probably to the extinction by circumstellar dust, just like the C-rich Miras. Indeed, some of them are spectroscopically confirmed to have circumstellar dust (\citealt{groenewegen2009}). These dusty O-rich Miras with very long period may be counterparts to the Galactic OH/IR stars.

To derive the PMRs, we concentrate on the O-rich Miras (i.e., black dots in Figure~\ref{pl}). The scatter about the PMR is related to difference in the color, error in the photometry, difference in the observed pulsation phase, and its intrinsic error. Mira-like variables discussed here are bright especially in the infrared, and their photometric errors should be small. Meanwhile, they are large-amplitude variables, and the difference in the observed pulsation phase (recall that we use single-epoch photometric data) can produce significant scatter. With the present data, we can not estimate the attributable fraction of scatter for each possible cause. Then we assume that the primary cause of the scatter about the PMR is due to the difference in the observed pulsation phase, and derive the PMRs in the following way. As a first step, we calculate the mean apparent magnitude ($<m_i>$) of O-rich Miras in the $\log_{10} P$ (days) $= 0.05$ bin and its corresponding standard deviation ($\sigma_{m_i}$), where the suffix $i$ stands for the $i$-th bin. The cyan points with error bars in Figure~\ref{pl} show the calculated $<m_i>$ and $\sigma_{m_i}$. Because the cyan points obviously do not lie on a line, we fit two lines in the form of $m_\lambda = a_\lambda \log_{10} P + b_\lambda$ to the cyan points by using 
least-square algorithm. We gave a relative statistical weight for each data point, which is proportional to $1/\sigma_{m_i}^2$. Apparently, the break seems to occur at around $\log_{10} P$ (days) $= 2.6$. In the following section, we estimate the break periods quantitatively. The calculated PMRs are shown as the blue thick solid and dashed lines in Figure~\ref{pl}. The values of the $a_\lambda$ and $b_\lambda$ with their errors and the $\log_{10} P$ range used for the fitting are summarized in Table~\ref{plrtable}. Also, we calculated the residuals $r$ from the derived PMRs for all O-rich Miras, which is defined as $r = a_\lambda \times \log P + b_\lambda - m_{\textrm{observed}}$. The standard deviation of $r$ is then calculated after 10 times iterations of 3 sigma clipping. The standard deviation of $r$ and the number of O-rich Miras used to calculate it are also tabulated in Table~\ref{plrtable}. The PMRs of C-rich Miras are not derived because most of them (except optical ones with periods shorter than about 400 days) do not obey a relation in the optical and near-infrared wavebands. In addition, their PMRs in the \textit{Spitzer} wavebands have very large scatter, probably due to the broad range in the amount of circumstellar extinction and/or in the amplitude of light variation.

\subsection{A kink in the period-magnitude relation}
It is clear that the cyan points do not fall on a single straight line. There is a break at a certain pulsation period, and the PMR becomes steep from there. This feature is seen regardless of the wavebands. We determined the break period by calculating $\alpha_i = (<m_{i+1}> - <m_i>) / 0.05$, where the $\alpha_i$ should remain constant (equals to the slope) for a straight line without a kink.
Although the calculated $\alpha_i$ is a bit noisy, we found a significant leap of $\alpha_i$ at a certain pulsation period. The corresponding periods ($\log_{10} P$) for the leap are, 2.7 for $<I>, J, H, K_s$, and $m_{\textrm{bol}}$, 2.65 for $[3.6], [4.5], [5.8]$, and $[8.0]$, and 2.6 for $[24]$, respectively. 

\citet{feast1989} first pointed out the existence of O-rich Miras in the LMC with periods longer than 420 days ($\log_{10} P \sim$ 2.6) that locate \textit{above} the extrapolations of the period-$K$ magnitude relation derived with Miras with periods shorter than 420 days. Then \citet{hughes1990} and \citet{whitelock2003} increased the number of samples, showing that the period-$m_{\textrm{bol}}$ relation also have a break at around that period. Our results are very consistent with these previous works. A possible explanation for the excess luminosity of O-rich Miras with longer periods is the "hot bottom burning" (HBB) process (e.g., \citealt{whitelock2003}, \citealt{feast2009}) that occurs in stars with masses greater than about 3 M$_\odot$ with the metallicity of the LMC (e.g., \citealt{garcia2006}). Note that the lower mass limit to activate the HBB depends on the metallicity (see, for example, \citealt{boothroyd}). Very interestingly, all but one spectroscopically-known infrared C-rich Miras have periods $longer$ than the break period. Pulsation period of a Mira variable is related to its mass and radius (e.g., \citealt{wood1990}). Assuming that the radii of O-rich and C-rich Miras with about the same pulsation period are more or less the same, the longer period one should have larger mass. On the condition that the kink in the PMR is due to the HBB, one can speculate that the C-rich Miras with periods longer than the break period are undergoing the HBB process now. If so, the high mass-loss rate ($> 10^{-5}$ M$_\odot$/yr, see next section) of infrared C-rich Miras can be due to the excess luminosity emerged from the HBB process that should create additional radiation pressure to the dust grains. See, for example, \citet{marigo2007} for the theoretical view on the HBB process in C-rich AGB stars.

\begin{table}
\caption[]{Period-luminosity relations for O-rich Mira variables in the LMC of the form $m_{\lambda} = a_\lambda \times \log P + b_\lambda$.}
\label{plrtable}
\centering
\begin{tabular}{lrrrrr}
\hline
\multicolumn{1}{l}{Band} & \multicolumn{1}{c}{$a_\lambda$} & \multicolumn{1}{c}{$b_\lambda$} & \multicolumn{1}{c}{fit range$^a$} & \multicolumn{1}{c}{$\sigma$} & \multicolumn{1}{c}{N$^b$} \\
\hline
$I$                & $-$1.077$\pm$0.164  & 16.653$\pm$0.381 & [2.1:2.6]    & 0.283 & 372 \\
$J$                & $-$3.389$\pm$0.108  & 20.399$\pm$0.241 & [2.0:2.6]    & 0.261 & 395 \\
$J$                & $-$5.079$\pm$0.797  & 24.733$\pm$2.216 & [2.6:2.95]   & 0.398 &  49 \\
$H$                & $-$3.498$\pm$0.095  & 19.880$\pm$0.216 & [2.0:2.6]    & 0.272 & 395 \\
$H$                & $-$5.579$\pm$0.758  & 25.191$\pm$2.114 & [2.6:2.95]   & 0.443 &  48 \\
$K_s$              & $-$3.675$\pm$0.076  & 19.956$\pm$0.173 & [2.0:2.6]    & 0.238 & 396 \\
$K_s$              & $-$6.850$\pm$0.901  & 28.225$\pm$2.493 & [2.6:2.95]   & 0.360 &  47 \\
$[3.6]$            & $-$3.913$\pm$0.120  & 20.053$\pm$0.275 & [2.0:2.6]    & 0.168 & 378 \\
$[3.6]$            & $-$5.583$\pm$0.470  & 24.151$\pm$1.313 & [2.6:2.95]   & 0.304 &  55 \\
$[4.5]$            & $-$3.744$\pm$0.109  & 19.556$\pm$0.254 & [2.0:2.6]    & 0.187 & 374 \\
$[4.5]$            & $-$5.239$\pm$0.167  & 23.042$\pm$0.503 & [2.6:3.1]    & 0.331 &  64 \\
$[5.8]$            & $-$3.719$\pm$0.117  & 19.311$\pm$0.272 & [2.0:2.6]    & 0.209 & 379 \\
$[5.8]$            & $-$5.783$\pm$0.285  & 24.381$\pm$0.851 & [2.6:3.1]    & 0.307 &  64 \\
$[8.0]$            & $-$3.742$\pm$0.119  & 19.177$\pm$0.273 & [2.0:2.6]    & 0.220 & 377 \\
$[8.0]$            & $-$8.938$\pm$0.327  & 32.850$\pm$0.980 & [2.6:3.1]    & 0.384 &  64 \\
$[24]$             & $-$13.182$\pm$0.803 & 43.213$\pm$2.336 & [2.6:3.1]    & 0.623 &  62 \\
$m_{\textrm{bol}}$ & $-$3.321$\pm$0.211  & 21.963$\pm$0.495 & [2.2:2.6]    & 0.215 & 137 \\
$m_{\textrm{bol}}$ & $-$4.270$\pm$0.743  & 24.019$\pm$2.094 & [2.65:2.95]  & 0.275 &  41 \\
\hline
\end{tabular}
$~^a$ in $\log_{10} P$, $~^b$ Number of O-rich Miras used to calculate the standard deviation ($\sigma$) of residuals from the fit, where the residual is defined as : $a_\lambda \times \log P + b_\lambda - m_{\textrm{observed}}$.
\end{table}

\begin{figure}
\centering
\includegraphics[scale=0.34,angle=-90]{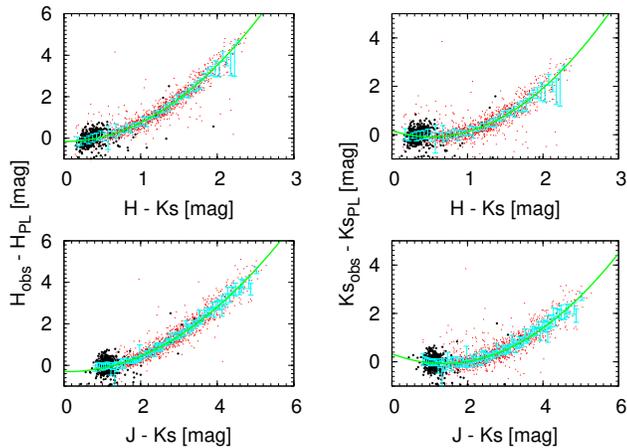}
\caption{The correlation between the observed near-infrared color of all color-classified O-rich (black dots) and C-rich (red dots) Miras and the "deviation" (see text). The solid line is the best-fit of a quadratic function in the form of $d = a x^2 + b x + c$, where the $d$ and $x$ are the deviation and the employed near-infrared color in magnitude. The fitting results are summarized in Table~\ref{devcolor}.}
\label{deviation}
\end{figure}

\begin{table}
\caption[]{Color-deviation relations for C-rich Mira variables in the LMC of the form deviation [mag] $= a x^2 + b x + c$, where $x$ is the employed color.}
\label{devcolor}
\centering
\begin{tabular}{lrrrr}
\hline
\multicolumn{1}{c}{deviation} & \multicolumn{1}{c}{$x$} & \multicolumn{1}{c}{$a$} & \multicolumn{1}{c}{$b$} & \multicolumn{1}{c}{$c$} \\
\hline
$H_{obs}-H_{PL}$                      & $H-K_s$  & 0.939$\pm$0.029 & $-$0.038$\pm$0.077 & $-$0.155$\pm$0.045 \\
$H_{obs}-H_{PL}$                      & $J-K_s$   & 0.205$\pm$0.010 & $-$0.041$\pm$0.056 & $-$0.301$\pm$0.071 \\
$K_{s_{obs}} - K_{s_{PL}}$        & $H-K_s$  & 0.924$\pm$0.029 & $-$0.968$\pm$0.076 &      0.166$\pm$0.044 \\
$K_{s_{obs}} - K_{s_{PL}}$        & $J-K_s$   & 0.213$\pm$0.009 & $-$0.586$\pm$0.052 &      0.331$\pm$0.066 \\
\hline
\end{tabular}
\end{table}

\subsection{Circumstellar extinction}
Figure~\ref{pl2} and Table~\ref{t-test} suggest that the optical O-rich and C-rich Miras fall on almost the same PMRs in optical and near-infrared wavebands, but infrared Miras do not. It is well known that the PMR is best defined at $K-$band, where the atmospheric bands do not contribute much and the fluxes correspond to the photosphere. It seems that this explanation holds for the optical Miras only. The $K_s-$band fluxes of the infrared Miras, however, do not represent their photospheres any more because of the heavy circumstellar extinction even in the $K_s-$ band. \cite{wood2003} studied MSX (\citealt{egan}) sources in the LMC, and found that infrared Mira-like variables do not fall on the standard period$-K_s$ magnitude relation for Mira-like variables (See, for example, sequence C in \citealt{wood2000} and \citealt{ita2004}), and they are fainter than the luminosity that the PMR predicts. Here we assume that the reason of the reduced optical and near-infrared flux of the infrared Miras is due solely to the circumstellar extinction by their surrounding dust shell. We also assume that all of the infrared Miras should have fallen on the extension of the optical and near-infrared PMR of optical Miras if they were without circumstellar dust. Under these assumptions, the deviation from the luminosity that the PMR predicts (i.e., $m_{\lambda, \textrm{deviation}} = m_{\lambda, \textrm{observed}} - m_{\lambda, \textrm{pmr}}$) should be a measure of the strength of circumstellar extinction (\citealt{matsunaga2005}, \citealt{ita2009}). It is interesting to note that there is no relevance between the deviation and pulsation period (see Figures~\ref{pl} and \ref{pl2}). It is also worth mentioning that the deviation is also independent of the pulsation amplitude at near-infrared bands. The interpretation of these observational facts will be made in the separate paper (Ita et al., in preparation). 

As stated above, the amount of deviation is independent of pulsation period and amplitude. However, it would be of great interest if we can correlate the deviation with the other observables (and especially distance-independent ones), because such relation would be useful, for example, to correctly estimate the distances to Mira-like variables through PMR after considering the reduction in their apparent magnitude by the circumstellar extinction. Figure~\ref{deviation} is a plot to show that there is a strong correlation between the observed near-infrared color and the deviation. This result suggests that the near-infrared color can be a good indicator to probe the circumstellar extinction of Miras. To formulate the relation between the near-infrared colors and the deviation, we calculate the mean deviation ($<d_i>$) of Miras in the $x = 0.05$ mag bin and its corresponding standard deviation ($\sigma_{d_i}$), where the suffix $i$ stands for the $i$-th bin and $x$ represents the employed color in magnitude. The cyan points with error bars in Figure~\ref{deviation} show the calculated $<d_i>$ and $\sigma_{d_i}$. Then we fit a quadratic function to the cyan points in the form of $d = a x^2 + b x + c$ by using Levenberg-Marquardt algorithm with giving statistical weights proportional to $1/\sigma_{d_i}^2$. The fitting results are shown in the solid line in Figure~\ref{deviation}, and the calculated coefficients are summarized in Table~\ref{devcolor}.

\begin{figure}
\centering
\includegraphics[scale=0.34,angle=-90]{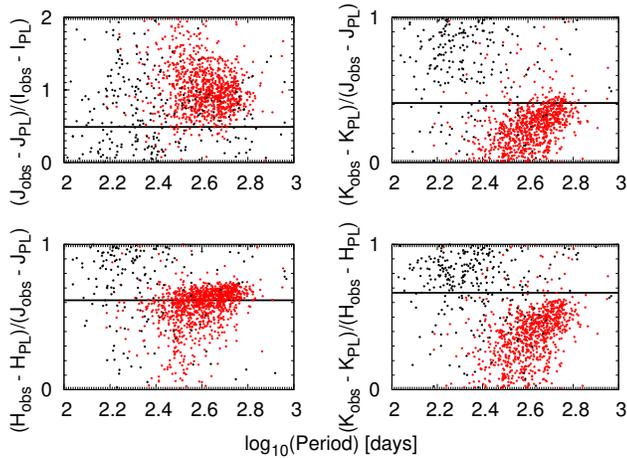}
\caption{The correlation between the pulsation period and the ratio of deviations at two different wavebands. Black and Red dots represent O-rich and C-rich Miras, respectively. The horizontal lines show the values expected by standard interstellar extinction law (\citealt{cardelli1989}).}
\label{deviationratio}
\end{figure}

The amount of deviation itself has important physical meanings. It can be used to roughly estimate the mass-loss rate by making a further assumption that the standard interstellar extinction law can be equally applied to Mira's circumstellar environment. This assumption should be inappropriate especially for C-rich Miras, because the interstellar environment is O-rich in the first place. Also, the extinction law in the circumstellar environment looks indeed different from that in the interstellar environment, to the extent that the Figure~\ref{deviationratio} shows (see next). At the level of order of magnitude estimate, however, we suppose the assumption is valid. The observed $K_s$ band magnitude of most of the infrared Miras are more than 2 magnitude fainter than the one predicted by the period$-K_s$ magnitude relation. There are some extreme infrared Miras that are more than 4 magnitude fainter than the prediction. Using the standard interstellar extinction law, the 2(4) magnitude extinction in $K_s$ band correspond to $A_\textrm{v}$ of about 17(34) mag, and the optical depth at visual wavelength of about 15.7(31.3). \citet{vanloon2007} related the total mass loss rate and the optical extinction as, $\dot{M} [M_\odot yr^{-1}] = 1.5 \times 10^{-9} (Z/Z_\odot)^{-0.5} (L/L_\odot)^{0.75} A_\textrm{v}^{0.75}$. Substituting typical LMC values of $Z=0.4 Z_\odot$ and $L=8000 L_\odot$, the circumstellar extinction of $A_{\textrm{v}}=15.7(31.3)$ mag gives $\dot{M} \sim 1.6(2.7) \times 10^{-5}$ $M_\odot yr^{-1}$. If we take $L=4000 L_\odot$, instead of $L=8000 L_\odot$,  the mass loss rate is reduced by a factor of (4000/8000)$^{0.75}$ $\sim$ 0.6. At any rate, it is likely that many of the infrared Miras in the LMC are losing mass of the order of about 10$^{-5}$ $M_\odot yr^{-1}$, comparable to the Galactic (i.e., presumably more metal-rich) Miras. Interestingly, this rather rough estimation of mass-loss rate is consistent with \cite{groenewegen2009}, who fit model spectral energy distribution to the observed mid-infrared spectra to calculate mass-loss rates of the same stars.

The ratios of deviations in any two different wavebands also give us insights on the properties (e.g., grain size and composition) of the circumstellar dust. The basic idea is like the interstellar extinction law that varies with the properties of interstellar dust. Figure~\ref{deviationratio} is a plot to show the relation between the ratios of deviations and pulsation period of Miras. For comparison purpose, the ratio of standard interstellar extinction law (\citealt{cardelli1989}) is also shown by the horizontal line. It is clear that the extinction law in the interstellar and circumstellar environment is different. It is also clear that the ratio of deviations (hence dust properties) changes with increasing pulsation period of C-rich Miras. Interestingly enough, the interstellar and circumstellar extinction law differs considerably for shorter period Miras, but the difference is little for longer period ones. It would be valuable to compare this observational results with the dust extinction models to quantitatively constrain the circumstellar dust properties.

\section{Conclusions}
We derived the near- to mid-infrared period-magnitude and also period-bolometric luminosity relations of OGLE-III Mira-like variables in the LMC. Mass-losing Mira-like variables do not fall on the derived relations, due to the circumstellar extinction. We showed that this fact can be used as an proxy to study the properties of the circumstellar dust, and also to estimate mass-loss rate. We found that the dust properties change with increasing pulsation period. A comparison between the observational facts and dust extinction model would reveal how dust grain evolves during the Mira phase.

\section*{Acknowledgments}
We thank Prof. Michael Feast, Prof. Yoshikazu Nakada, Dr. Toshihiko Tanab\'{e} for useful comments on the first manuscript of this paper. This work is supported by the Grant-in-Aid for Encouragement of Young Scientists (B) No.~21740142 from the Ministry of Education, Culture, Sports, Science and Technology of Japan. This publication makes use of data products from the Two Micron All Sky Survey, which is a joint project of the University of Massachusetts and the Infrared Processing and Analysis Center/California Institute of Technology, funded by the National Aeronautics and Space Administration and the National Science Foundation.

\bsp

\label{lastpage}

\end{document}